\documentclass[aps,amsfonts,amsmath,superscriptaddress,twocolumn,showpacs,floatfix,nofootinbib,showkeys]{revtex4-1}

\usepackage{marginnote}
\usepackage{comment}
\usepackage{epsfig}
\usepackage{graphicx}
\usepackage{mathptmx}      
\usepackage{latexsym}
\usepackage{bm}
\usepackage{bbold}
\usepackage{color}

\usepackage{stackrel}

\def\slashchar#1{\setbox0=\hbox{$#1$}
   \dimen0=\wd0 \setbox1=\hbox{/} \dimen1=\wd1
   \ifdim\dimen0>\dimen1 \rlap{\hbox to \dimen0{\hfil/\hfil}} #1
   \else  \rlap{\hbox to \dimen1{\hfil$#1$\hfil}} / \fi}

\newcommand{\ignore}[1]{}

\newcommand{\x}{\mathbf{x}}

\newcommand{\gap}{\sigma}

\newcommand{\brane}{b}

\begin{document}

\title{Cosmological Dark Matter from a Bulk Black Hole}
\author{Sylvain Fichet}
\email{sfichet@caltech.edu}

\affiliation{ICTP South American Institute for Fundamental Research  \& IFT-UNESP, R. Dr. Bento Teobaldo Ferraz 271, S\~ao Paulo, Brazil}

\affiliation{Centro de Ciencias Naturais e Humanas (CCNH), Universidade Federal do ABC,  
Santo Andre, 09210-580 SP, Brazil}

 \author{Eugenio Meg\'{\i}as}
 \email{emegias@ugr.es}

 \affiliation{Departamento de F{\'\i}sica At\'omica, Molecular y Nuclear and
   Instituto Carlos I de F{\'\i}sica Te\'orica y Computacional, \\ Universidad
   de Granada, Avenida de Fuente Nueva s/n, E-18071 Granada, Spain.}

\author{Mariano Quir\'os}
\email{quiros@ifae.es}

\affiliation{Institut de F\'{\i}sica d'Altes Energies (IFAE) and The Barcelona Institute of  Science and Technology (BIST), Campus UAB, 08193 Bellaterra (Barcelona) Spain}

\date{\today}

\begin{abstract}
  \medskip

We study the cosmology of a three-brane in a specific five-dimensional scalar-gravity (\textit{i.e.}~soft-wall) background, known as the linear dilaton background. 
We discover that the Friedmann equation of the brane-world automatically contains {a term mimicking pressureless matter. We propose to identify this  term as dark matter.}
This dark matter arises as a projection of the bulk black hole on the brane, which contributes to the  brane Friedmann equation via  both the Weyl tensor and the scalar stress tensor. 
The nontrivial matter-like behavior  is due to an exact cancellation between the Weyl and scalar pressures.  
 We  show that the Newtonian potential only receives a mild short-distance correction going as { inverse distance squared},  ensuring  compatibility of the linear dilaton brane-world with observed 4D gravity.
Our setup can be viewed as a consistent cosmological description of the
 holographic theories arising in the  linear dilaton background. 
{We also present more general scalar-gravity models where the brane cosmology features an effective energy  density whose behavior smoothly interpolates between dark radiation, dark matter and dark energy depending on a model parameter.}

\end{abstract}



\maketitle

\raggedbottom

\section{Introduction}

Five-dimensional (5D) gravity coupled to a scalar field has proven to be a fecund playground, leading to a host of theoretical results  and  models of the real world  (see \textit{e.g.}~\cite{Karch:2006pv,Gursoy:2007cb,Gursoy:2007er,Gubser:2008ny,Falkowski:2008fz, Batell:2008zm, Batell:2008me,Cabrer:2009we,vonGersdorff:2010ht, Cabrer:2011fb,Megias:2019vdb, Megias:2021mgj,Megias:2021arn}).
Our focus in this letter is a specific 5D scalar-gravity background (\textit{i.e.}~a soft-wall background) which is  sometimes  referred to as the ``linear dilaton background’’.  This model is known to have peculiar thermodynamic~\cite{Gursoy:2007cb} and field theoretical properties~\cite{Cabrer:2009we,Megias:2019vdb,Megias:2021mgj,Megias:2021arn}. For example,  all quantum fields living on the linear dilaton background have a spectral distribution    that features a gapped continuum. 
This feature  has been recently used in extensions  of the Standard Model~\cite{Csaki:2021gfm,Csaki:2021xpy}. 
\\

In this work we put the linear dilaton background at finite temperature and posit a flat 3-brane moving over the  background, in the spirit of brane-world models (see \textit{e.g.}~\cite{Brax:2003fv}).  We discover  a surprising property: from  the viewpoint of a brane observer, the local  Friedmann equation automatically contains {an effective energy term that may be identified as}  \textit{dark matter}.
This dark matter emerges as a nontrivial effect from the bulk physics projected on the brane. It originates from a combination of the 5D Weyl tensor and of the bulk scalar vev, as we will demonstrate further below.  
\\
 
To bring this result into context, we remind that there is a notorious  analog  in pure Anti-de Sitter (AdS) background, that has been gradually uncovered and studied in~\cite{  Shiromizu:1999wj,Binetruy:1999hy,Hebecker:2001nv,Langlois:2002ke,Langlois:2003zb}. In pure AdS the net effect of the bulk physics projected on the brane  gives rise to {\textit{radiation}}, {which is identified as cosmological dark radiation in the context of a brane-world. }
This remarkable fact is in direct connection with the fact that the bulk black hole in AdS-Schwarzschild background corresponds to the thermal state in the holographic CFT, perhaps one of the most fascinating
entries of the AdS/CFT correspondence \cite{Aharony:1999ti}. 
In our case, by performing the analogous calculation  with the linear dilaton background,  we discover  that the bulk black hole gives rise to {dark \textit{matter}}.
\\

Decades of astronomical observations point to the existence of dark matter. Determining its  nature is a pressing question in fundamental physics. While a common hypothesis is that dark matter may be a new particle (that remains so far elusive), our study leads to a fundamentally different viewpoint. Our setup provides, in a sense, an origin to cosmological dark matter via a modification of gravity. 
See \textit{e.g.}~\cite{Clifton:2011jh} for  a few other attempts to explain dark matter via modified gravity. 
\\

In this paper we present the derivation of our central result, the effective Friedmann  equation of the linear dilaton brane-world, that is shown to contain dark matter. We present  nontrivial consistency checks of this result.
We also compute the deviation to the Newtonian potential. { We then outline more general  models featuring a variety of equations of state depending on a model parameter,} and discuss some conceptual points and prospects. Extra developments and technical details are laid out 
in~\cite{Fichet:2022ixi}, which can be considered as a companion to this letter.

\section{The 5D scalar-gravity system}

We consider the general scalar-gravity action in the presence of a brane,
\begin{eqnarray}
S &=& \int d^5x \sqrt{g} \left( \frac{M_5^3}{2} {}^{(5)}R - \frac{1}{2} (\partial_M \phi)^2 - V(\phi)   \right) \nonumber \\
&-& \int_{\textrm{brane}} d^4x \sqrt{\bar g}\, (V_{\brane}(\phi) + \Lambda_{\brane} ) + \ldots  \, \label{eq:action}
\end{eqnarray}
 ${}^{(5)}R$ is the 5D Ricci scalar, $\phi$ is the scalar field, $M_5$ is  the fundamental 5D Planck scale,
 $\bar g_{\mu\nu}$ is  the induced metric on the brane,
$g \equiv |\det g_{MN}|$ and $\bar g \equiv |\det \bar g_{\mu\nu}|$ are the metrics determinants,  $\Lambda_b$ is the brane tension,  $V$ and  $V_b$ are the bulk and brane-localized potentials for $\phi$. We assume that the brane potential sets the scalar field vacuum expectation value (vev) to a nonzero value $\langle\phi\rangle=v_{\brane}$, with $V_{\brane}(v_{\brane})=0$.
The bulk potential is explicitly given further below.
The ellipses encode the Gibbons-Hawking-York term~\cite{York:1972sj,Gibbons:1976ue} and the action of quantum fields living on the 5D background. 

The 5D metric is written in a frame suitable for brane cosmology  as
\begin{align}
ds^2 & = g_{MN} dx^M dx^N \equiv -n^2(r)d\tau^2 + \frac{r^2}{\ell^2} d\x^2 + b^2(r) dr^2 \label{eq:brane_frame}
\,.
\end{align}
We allow the existence of a black hole horizon encoded in the $n(r)$ and $b(r)$ factors, the position of the horizon being given by $n(r_h) = 0 = 1/b(r_h)$. Latin indices $(M,N,\cdots)$  refer to 5D coordinates,  Greek indices $(\mu,\nu,\cdots)$ refer to 4D coordinates.
 
\begin{figure}[t]
\centering
\includegraphics[trim={5cm 2cm 4cm 2cm},clip,width=0.48\textwidth]{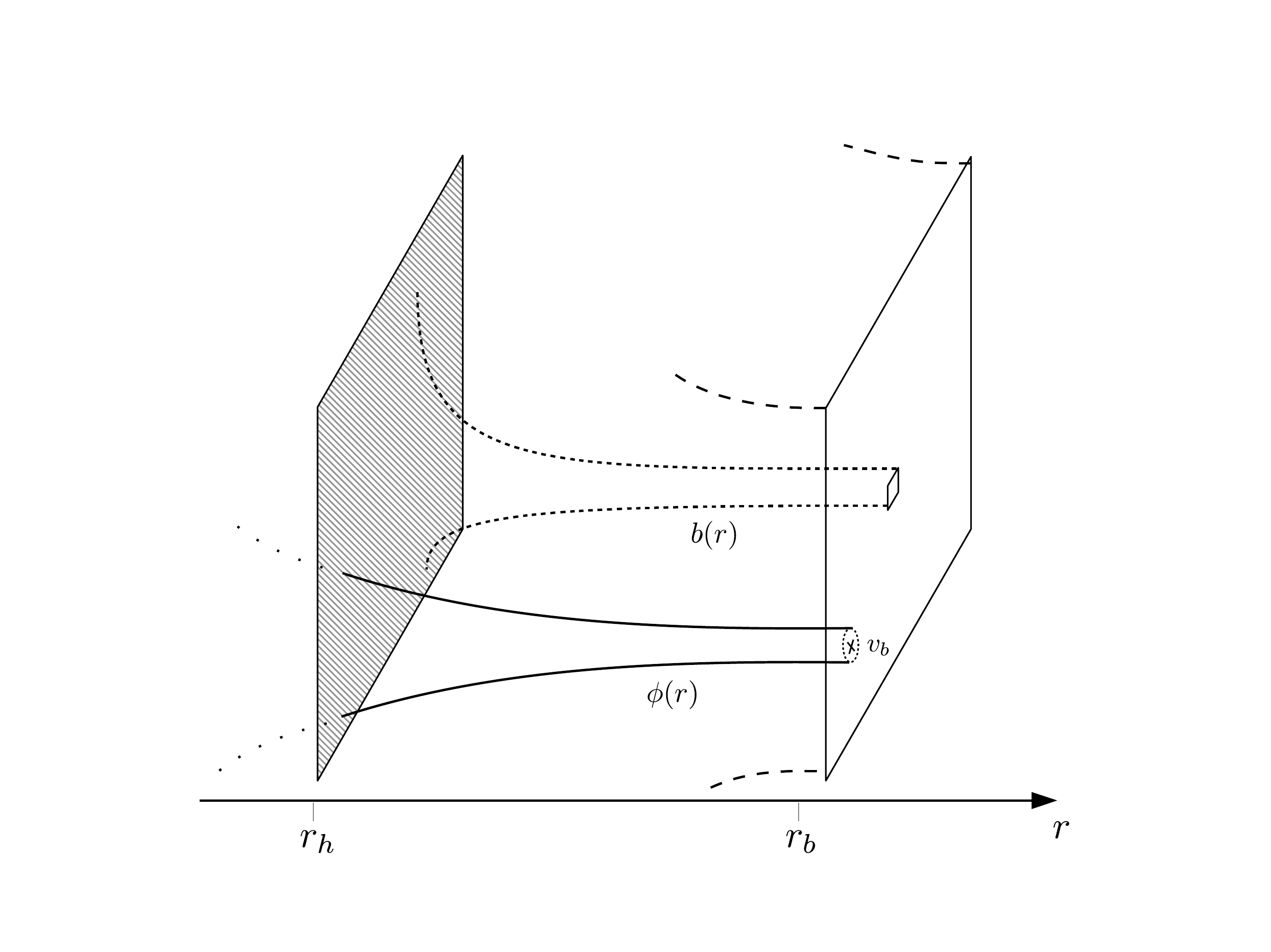}
\caption{\it 
Overview of the scalar-gravity system.  The brane and the  horizon are located respectively at $r=r_b$ and $r=r_h$.  The  scalar vacuum expectation value is fixed by a brane potential to a value $v_b$ which remains constant when $r_b$ varies.   The  scalar VEV (plain) evolves in the bulk, 
the  blackening factor of the metric (dotted) diverges at the horizon. 
}
\label{fig:cartoon}
\end{figure}

The 3-brane is localized at the position $r = r_{\brane}$. 
Our frame \eqref{eq:brane_frame} is appropriate to describe  cosmology as seen from the brane standpoint. The induced metric  on the brane is 
\begin{align}
ds^2 & = \bar g_{\mu\nu} dx^\mu dx^\nu  
 \equiv -dt^2 + \frac{r_{\brane}^2}{\ell^2} d\x^2 \label{eq:brane_induced}
\,,
\end{align}
where we have introduced the brane cosmic  time $dt=n(r_{\brane})d\tau$. 
According to this metric, if the brane moves along $r$ in the 5D background,  the observer perceives expansion of the four-dimensional (4D) universe  with Hubble parameter $H=\dot{r_b}/r_b$, where $\dot{r_b}\equiv \partial_t r_b$.  
We choose that $r_b$ equals $\ell$ at present times, such that  $r_{\brane} = a(t)\ell$ where $a(t)$ is the standard scale factor.   {An overview of the brane-world is shown in Fig.~\ref{fig:cartoon}.}

The 5D equations of motion of the system are 
\begin{align}
{}^{(5)}G_{MN}=\frac{1}{M_5^3}T^\phi_{MN}\,,\quad 
\frac{1}{\sqrt{g}} \partial_M\left(\sqrt{g} g^{MN}\partial_N  \phi \right) = \frac{\partial V}{\partial \phi} \,,
\label{eq:GMN}
\end{align}
with  ${}^{(5)}G_{MN}={}^{(5)}R_{MN} - \frac{1}{2} g_{MN} {}^{(5)}R $ and $T^\phi_{MN}= 
\partial_M \phi  \partial_N \phi -
 g_{MN} \left[ \frac{1}{2}(\partial_A\phi)^2 +  V(\phi)\right]  $. 
More explicitly,  the equations of motion for the 5D background in the cosmological frame  are~\cite{Megias:2018sxv} 
\begin{eqnarray}
\hspace{-0.3cm}&& \frac{n^{\prime\prime}(r)}{n(r)} - \left(\frac{n^\prime(r)}{n(r)} - \frac{1}{r}  \right) \left( \frac{b^\prime(r)}{b(r)} - \frac{2}{r} \right) = 0 \,,  \label{eq:EoM1}  \\
\hspace{-0.3cm}&&\frac{n^\prime(r)}{n(r)} + \frac{b^\prime(r)}{b(r)} - r \bar\phi^\prime(r)^2 = 0 \,,   \nonumber
\\
\hspace{-0.3cm}&&\frac{n^\prime(r)}{n(r)} + \frac{1}{r} +  r\, b^2(r) \bar V(\bar\phi) - \frac{r}{2}\bar\phi^\prime(r)^2 = 0 \,,    \nonumber
\\
\hspace{-0.3cm}&& \bar\phi^{\prime\prime}(r) + \left( \frac{n^\prime(r)}{n(r)} - \frac{b^\prime(r)}{b(r)} + \frac{3}{r} \right) \bar\phi^\prime(r) -  b^2(r) \frac{\partial \bar V}{\partial \bar\phi} = 0 \,,  \nonumber
\end{eqnarray}
with the dimensionless  field~$\bar\phi \equiv \phi / (3 M_5^3)^{1/2}$ and $\bar V\equiv V/(3 M_5^3)$. 
Importantly, even though one of these differential equations seems redundant, it cannot be ignored because it still implies a nontrivial \textit{algebraic} relation between the integration constants. Also notice that the integration constants can depend on $r_b$ through the boundary conditions; the brane location thus influences the 5D background. 

We turn to gravity from the brane viewpoint. 
The effective 4D Einstein equation seen by a brane observer  is computed from the 5D Einstein equation by projecting on the brane via the Gauss equation together with the Israel junction condition\,\cite{Shiromizu:1999wj}. Introducing the unit vector $n_M$ normal to the brane  that satisfies  $n_M n^M = 1$ and  $\bar g_{MN} = g_{MN} - n_M n_N$, the 4D Einstein equation on the brane is
\begin{eqnarray}
{}^{(4)}G_{\mu\nu} = 
\frac{1}{M^2_{\rm Pl}}\left(T_{\mu\nu}^b+ T^{\rm eff}_{\mu\nu}\right)
+O\left(\frac{T_b^2}{M^6_5}\right) \label{eq:EE_brane}
\end{eqnarray}
with  ${}^{(4)}G_{\mu\nu} = {}^{(4)}R_{\mu\nu} - \frac{1}{2} \bar g_{\mu\nu} {}^{(4)}R$, and $T^b_{\mu\nu}$ the stress tensor of brane-localized   matter. 
The { ``holographic''} effective stress tensor $T^{\rm eff}_{\mu\nu}=  \tau^{W}_{\mu\nu}+\tau^{\phi}_{\mu\nu}+\tau^{\Lambda}_{\mu\nu}$  contains: 

\textit{i)} The projection of the 5D Weyl tensor  ${}^{(5)}C^{M}{}_{NPQ}$ on the brane
\begin{align}
\frac{1}{M^2_{\rm Pl}} \tau^{W}_{\mu\nu} = - {}^{(5)}C^{M}{}_{NPQ} n_M n^P \bar g_\mu{}^{N} \bar g_\nu{}^Q \,,
\end{align}
leading to corresponding values of the energy density $\rho^W$ and pressure $P^W$ given by
\begin{align}
    \rho^W&=\frac{3}{2}\frac{M^2_{Pl}}{b^2(r_b)r_b}\left(\frac{n^\prime(r_b)}{n(r_b)}-\frac{1}{r_b}  \right) \,, \nonumber\\
    P^W&=\frac{1}{2}\frac{M^2_{Pl}}{b^2(r_b)r_b}\left(\frac{n^\prime(r_b)}{n(r_b)}-\frac{1}{r_b}  \right) \,,
    \label{eq:rhoPW}
\end{align}
where we have made use of Eqs.~(\ref{eq:EoM1}).

\textit{ii)} The projection of the bulk stress tensor
\begin{align}
\frac{1}{M^2_{\rm Pl}} \tau^{\phi}_{\mu\nu}&=\frac{2}{3 M_5^3} \left[ T^\phi_{MN} \bar g_{\mu}{}^{M} \bar g_{\nu}{}^{N} + \big(T^\phi_{MN} n^M n^N - \frac{1}{4} T^{\phi,M}_{M} \big) \bar g_{\mu\nu} \right] \nonumber\\
&=\frac{3}{2}\left(\frac{\bar\phi^{\prime}(r_b)^2}{2 b^2(r_b)} -\bar V \right)\bar g_{\mu\nu}  \,,
\end{align}
leading to the values of $\rho^\phi$ and $P^\phi$, after using the EoM (\ref{eq:EoM1}), 
\begin{align}
    \rho^\phi&=-\frac{3}{2}\frac{M^2_{Pl}}{b^2(r_b)r_b}\left(\frac{n^\prime(r_b)}{n(r_b)}+\frac{1}{r_b}  \right)  \,, \nonumber\\
    P^\phi &=\frac{3}{2}\frac{M^2_{Pl}}{b^2(r_b)r_b}\left(\frac{n^\prime(r_b)}{n(r_b)}+\frac{1}{r_b}  \right) \,.
    \label{eq:rhoPphi}
\end{align}
%

\textit{iii)} The contribution from the brane tension
\begin{align}
\frac{1}{M^2_{\rm Pl}} \tau^{\Lambda}_{\mu\nu}=-\frac{\Lambda_b^2}{12M_5^6} \bar g_{\mu\nu}\,,
\end{align}
which yields the values of $\rho^\Lambda$ and $P^\Lambda$ as 
\begin{equation}
    \rho^\Lambda=-P^\Lambda=\frac{M^2_{\rm Pl}\Lambda_b^2}{12 M_5^6}\,.
    \label{eq:rhoPLambda}
\end{equation}
The brane tension is ultimately tuned to set the  effective 4D cosmological constant to zero. 

We work in the low-energy regime  
\begin{equation}
|T_{\mu\nu}^b|\ll \frac{M_5^6}{M_{\rm Pl}^2} \label{eq:smallT}
\end{equation}
 which justifies neglecting the higher order terms in 
 \eqref{eq:EE_brane}. 
 This restriction  implies further simplifications below.

\section{Dark matter from the linear dilaton black hole}

The linear dilaton background is defined by the bulk (super)potential 
\begin{equation}
\bar W(\bar \phi)=\frac{2}{\ell}e^{\bar\phi},\quad   \bar V(\bar\phi)= -\frac{3}{2\ell^2}e^{2\bar\phi}\,. \label{eq:V_LD}
\end{equation}
Solving the equations of motion \eqref{eq:EoM1} with the potential \eqref{eq:V_LD} we find for the 5D background
\begin{eqnarray}
n(r) &=& \frac{r}{\ell} \sqrt{1- \frac{r_h^3}{r^3}} \,,  \label{eq:n_LD}\\
b(r) &=& \frac{\ell}{r_{\brane}} \frac{e^{-\bar v_{\brane}}}{\sqrt{1 - \frac{r_h^3}{r^3} }} \,,  \label{eq:b_LD}\\
\bar\phi(r) &=& \bar v_{\brane} - \log\left( \frac{r}{r_{\brane}} \right) \,, \label{eq:phi_LD}
\end{eqnarray}
where $r_h$ (an integration constant) is the location of the black hole horizon in the brane cosmology frame. 
The domain of the variable $r$ is the interval $[0,\ell]$, where $r=0$ is the metric singularity and $r=\ell$ is the value of the brane location today, while $0\leq r_h\leq r_b\leq \ell$.
Importantly, we can  notice that a power of $3$ appears in the Schwarzschild factors, in contrast with pure AdS$_5$ where, instead, there would be a power of $4$. 

We then evaluate the brane effective Einstein equation by plugging the bulk solutions into Eq.\,\eqref{eq:EE_brane}, and deduce the Friedmann equation. The mass scale that naturally appears in the physical quantities is 
\begin{equation}
\eta = \frac{M_5^3}{M^2_{\rm Pl}}  = \frac{e^{\bar v_{\brane}}}{\ell}\,. 
\end{equation}
Using the low-energy assumption \eqref{eq:smallT} 
 which here becomes
 \begin{equation}
 \rho_b\ll \eta^2M^2_{\rm  Pl} \label{eq:rho_small}
\end{equation}
we obtain the first Friedmann equation on the brane,
\begin{equation}
3 M^2_{\rm Pl} \left(\frac{\dot r_{\brane}}{r_{\brane}}\right)^2 = \rho_b +\rho_{\rm eff}+O\left(\frac{\rho_b^2} {\eta^2 M^2_{\rm Pl}} \right)\, \label{eq:EFE}
\end{equation}
with
\begin{equation}
    \rho_{\rm eff} =  3 \eta^2 M_{\rm Pl}^2 
    \frac{r^3_h}{r^3_{\brane}}  \,.
    \label{eq:rhoeff}
\end{equation}
The $\rho_{\rm eff}$ energy density term is the critical result. It is a nontrivial effect from the bulk physics: a combination of the  Weyl tensor and of the  scalar stress tensor contributions. This holographically-induced $\rho_{\rm eff}$ scales as $r_{\brane}^{-3}$, therefore it behaves as a nonrelativistic matter term in the 4D Friedmann equation. 
(The analogous calculation in AdS would instead give a $r^{-4}_{\brane}$ scaling, \textit{i.e.}~radiation).  

In the brane-world paradigm, we  identify the Standard Model fields as  brane-localized modes that give rise to the brane energy density $\rho_b$. The effective energy density $\rho_{\rm eff}$ in Eq.~(\ref{eq:EFE})  is then naturally identified as the dark matter energy density. In other words, the linear dilaton brane-world   automatically features dark matter.

From the expression of $\rho_{\rm eff}$ in Eq.~(\ref{eq:rhoeff}), the fraction of dark matter energy  in the Universe $\Omega_{\rm DM}=\rho_{\rm DM}/\rho_{\rm crit}$ (with $\rho_{\rm crit}= 3 H^2M^2_{\rm Pl}$) induced by the linear dilaton background  is then
\begin{equation}
\Omega_{\rm DM} = \left(\frac{\eta}{H}\right)^2 \left(\frac{r_h}{r_b}\right)^3\,.
\end{equation}
At present times we have $r_b=\ell$,  $\Omega_{{\rm DM},0}=0.26$ and 
$H_0=1.47 \times 10^{-42}\,\textrm{GeV}$.  
This provides a constraint between the model parameters given by
\begin{equation}
    r_h\simeq 0.64\ell\left( \frac{H_0}{\eta} \right)^{2/3} \,.
    \label{eq:rh}
\end{equation}
As in the Standard Cosmology, this dark matter  dominates the universe for temperatures $T\lesssim 0.7\,\textrm{eV}$ and is subdominant with respect to  radiation for higher temperatures.

The origin of the $r^{-3}_{\brane}$ scaling is better understood as follows.  
The effective energy density and pressure, which appear in the Friedmann and continuity equations, are defined as
\begin{equation}
\rho_{\rm eff}=\rho^W+\rho^\phi+\rho^\Lambda,\quad P_{\rm eff}=P^W+P^\phi+P^\Lambda \,,
\end{equation}
where $\rho^W$ and $P^W$ are given by Eq.~(\ref{eq:rhoPW}), $\rho^\phi$ and $P^\phi$ by Eq.~(\ref{eq:rhoPphi}), and $\rho^\Lambda$ and $P^\Lambda$, after imposing the condition for cancellation of the cosmological constant, $\Lambda_b=6\eta^2 M^2_{\rm Pl}$, are given by
\begin{equation}
    \rho^\Lambda=-P^\Lambda=3 \eta^2 M^2_{\rm Pl} \,.
    \label{eq:rhoPLambda0}
\end{equation}

A straightforward application of the BH solutions (\ref{eq:n_LD}) and (\ref{eq:b_LD}) yields
\begin{equation}
    \rho^W+\rho^\phi=-3\eta^2 M^2_{\rm Pl}\left(1-\frac{r_h^3}{r_b^3} \right)  \,,
\end{equation}
which combined with $\rho^\Lambda$ from (\ref{eq:rhoPLambda0}), yields the result which appears in Eq.~(\ref{eq:rhoeff}).

On the other hand, for the effective pressure $P_{\rm eff}$ using again Eqs.~(\ref{eq:n_LD}) and (\ref{eq:b_LD}) we get
\begin{equation}
    P^W+P^\phi=\frac{2 M^2_{\rm Pl}}{b^2(r_b)r_b}\left( \frac{n^\prime(r_b)}{n(r_b)}+\frac{1}{r_b} \right)=3\eta^2 M^2_{\rm Pl}  \,,
\end{equation}
which combined with Eq.~(\ref{eq:rhoPLambda0}) yields $P_{\rm eff}=0$, leading to the equation of state $w_{\rm eff}=P_{\rm eff}/\rho_{\rm eff}=0$

This explains the $r^{-3}_{\brane}$ scaling and  ensures that the 4D conservation equation \textit{i.e.}~that the 4D Bianchi identity $D^\mu{}^{(4)}G_{\mu\nu}=0$ is satisfied.  
The cancellation we report here is nontrivial, 
as it is unclear if there exists a symmetry that enforces it. 

Another nontrivial consistency check is at the level of the 5D conservation equation projected on the brane, which  takes the general form~\cite{Tanaka:2003eg,Langlois:2003zb}
\begin{equation}
\dot \rho_{\rm eff}
 +4H\rho_{\rm eff}+H T_{\,\, \mu}^{{\rm eff}\, \mu}= -2T_{MN}u^Mn^N\,.
 \label{eq:continuity}
\end{equation}
Notice the $4H$ factor arising due to 5D spacetime. 
On the  rhs, $n^N$ is the unit vector normal to the brane and  outward-pointing, and $u^M$ is the brane velocity vector satisfying $u_M u^M=-1$~\cite{Tanaka:2003eg,Langlois:2003zb}. In the low-energy regime we have
\begin{equation}
u^M\approx\left(\frac{1}{n},{\bm 0},H r_b\right) \,, \,\quad n^M \approx \left(H r_b \frac{b}{n},  {\bm 0} , \frac{1}{b}   \right) \,
\end{equation}
up to $ O\left(H^2/\eta^2\right)$. 
Using the explicit expression of $T_{MN}$ obtained from our scalar-gravity solutions,   Eqs.~\eqref{eq:n_LD}--\eqref{eq:phi_LD}, it turns out that $T_{MN}u^Mn^N=0$  in the low-energy regime. The calculation involves again beautiful cancellations, and it is detailed  in \cite{Fichet:2022ixi}.
One can then easily verify that the 5D conservation equation is satisfied by the effective energy density \eqref{eq:rhoeff}, ensuring that the framework is fully consistent.

The low-energy  regime Eq.\,\eqref{eq:rho_small} implies 
$H\ll \eta$ since the total energy density is $\rho\sim H^2 M^2_{\rm Pl} $;   
it is the only assumption made throughout the calculations. We worked at first order in $H/\eta$. The cancellations  observed in $T_{\mu\nu}^{\rm eff}$ and in the 5D conservation equation   occur up to small $ O(H^2/\eta^2)$ factors. 

\section{The Newtonian potential}

The Newtonian potential for the LD model at present times can be deduced from the graviton brane-to-brane propagator $G_{\bm 2}$ using the optical theorem \cite{Fichet:2022ixi}. We find the discontinuity of this propagator to be 
\begin{equation}
\text{Disc}_{s}[G_{\bm 2}(\sqrt{s})]=2\pi \delta(s) +
\frac{\sqrt{\frac{s}{\gap^2}-1}}{s}\theta\Big(s\geq\gap^2\Big) \,, \label{eq:DicG2}
\end{equation}
where $\sigma=3\eta/2$ is the mass gap. 
The $\delta$ term corresponds to the 4D graviton. The second term, which encodes the rest of the 5D graviton  fluctuations, forms a  \textit{gapped continuum} characteristic of the linear dilaton background \cite{Megias:2021mgj}. From this discontinuity we deduce that the Newtonian potential of the linear dilaton brane-world is
\begin{equation}
V_N(R)= -\frac{m_1 m_2}{M^2_{\rm Pl}\,R}\bigg(1+\Delta(R)\bigg) \,,
\end{equation}
with
\begin{equation}
\Delta(R) 
\approx \begin{cases}\frac{4}{3\pi \gap R}~{\rm if}~R\ll \frac{1}{\gap}   \\ O\Big(e^{-\gap R}\Big)~{\rm if}~R\gg \frac{1}{\gap}  
\end{cases} \,. \label{eq:VN_LD}
\end{equation}
We see that the deviation from the Newtonian potential  appears essentially below the distance scale $1/\gap$ corresponding to the inverse mass gap. The deviation to the potential goes as $\propto 1/R^2$, unlike the AdS case, where it goes as $1/R^3$. Micron-scale fifth force experiments~such as~\cite{Smullin:2005iv} mildly constrain the $\sigma$ scale  as $\sigma \gtrsim 10$\,meV. This constraint, along with Eq.~(\ref{eq:rh}), translates into an upper bound on the location of the bulk black hole horizon,  $r_h\lesssim 2.3\times 10^{-21}\ell$.

\section{Extensions and uniqueness of the  linear dilaton brane-world}

In the previous sections we have seen that the bulk black hole from the LD braneworld model characterized by the exponential potential Eq.~(\ref{eq:V_LD}) 
leads to a pressureless matter term on the brane. We may wonder whether such a behavior of $\rho_{\rm eff}$ is specific to the LD model or if it appears in other 5D scalar-gravity solutions. In the next subsections we provide hints of uniqueness by extending the model in two different directions.  We consider a model with an exponential superpotential (like that of the LD model) but with a different exponent, and a model where a constant is added to the exponential superpotential. In both cases the scalar-gravity solutions will depend on a parameter which reproduces the LD model for particular values, but generalizes it. These more general scalar-gravity solutions  are interesting per se. We leave an extended investigation for future work. Our focus here is mostly on  illustrating the uniqueness of the behavior of $\rho_{\rm eff}$ in the LD model.

\subsection{A generalized exponential potential}
\label{sec:extension}

In this section we generalize the (super)potential of the LD model given by Eq.~(\ref{eq:V_LD})
to 
\begin{equation}
    \bar W(\bar\phi)=\frac{2}{\ell}e^{\nu\bar\phi},\quad \bar V(\bar\phi)=-\frac{4-\nu^2}{2\ell^2}e^{2\nu\bar\phi} \,,
\end{equation}
where the LD model is reproduced for the value $\nu=1$, while the AdS model is reproduced for the value $\nu=0$. 

The solution to the 5D equations of motion (\ref{eq:EoM1}) is given by
\begin{align}
    n(r)&=\frac{r}{\ell}\sqrt{1-\left(\frac{r_h}{r} \right)^{4-\nu^2}} \,, \label{eq:n_extended}\\
    b(r)&=\left(\frac{r}{\ell} \right)^{\nu^2-1}  \left(\frac{\ell}{r_b} \right)^{\nu^2}\frac{e^{-\nu \bar v_b}}{\sqrt{1-\left( \frac{r_h}{r}\right)^{4-\nu^2}} }  \,, \label{eq:b_extended}\\
    \bar\phi(r)&=\bar v_b-\nu \log\left( \frac{r}{r_b} \right)  \,,
\end{align}
and the relation between the 5D and 4D Planck scales is given by
\begin{equation}
    M_5^3=\frac{1}{2}\bar W_b M_{\rm Pl}^2=\eta M_{\rm Pl}^2,\quad \eta\equiv\frac{1}{\ell} e^{\nu\bar v_b}\,.
\end{equation}

After using the relation for vanishing of the cosmological constant $\Lambda_b=3 M_5^3 \bar W_b=6\eta M_5^3$, one readily gets the brane vacuum energy and pressure as
\begin{equation}
    \rho^\Lambda=-P^\Lambda=3\eta^2 M_{\rm Pl}^2\, .
\end{equation}
Using the solution (\ref{eq:n_extended})-(\ref{eq:b_extended}) one easily gets
\begin{equation}
     \rho_{\rm eff}=3\eta^2 M_{\rm Pl}^2 \frac{r_h^{4-\nu^2}}{r_b^{4-\nu^2}},\quad 
     P_{\rm eff}=\eta^2 M_{\rm Pl}^2 (1-\nu^2) \frac{r_h^{4-\nu^2}}{r_b^{4-\nu^2}} \,,
\end{equation}
which yields an equation of state
\begin{equation}
    w_{\rm eff}=\frac{1-\nu^2}{3} \,.
\end{equation}
We can see that the dark matter behavior ($w_{\rm eff}=0$) appears only for $\nu=1$. 

Interestingly, the ``holographic'' effective energy density in this model interpolates from dark radiation behavior ($w_{\rm eff}=1/3$) for $\nu=0$ to dark energy  behavior ($w_{\rm eff}=-1$) for $\nu=2$. {For $0\le \nu \leq 2$ the solution satisfies the continuity equation  automatically, cf. Eq.~(\ref{eq:continuity}).} Finally, let us point out that the singularity at $r=0$ is a good one for $\nu\leq 2$~\cite{Cabrer:2009we}.~\footnote{{For $\nu=2$ the solution to the 5D equations of motion (\ref{eq:EoM1}) is $n(r) = r/\ell$, $b(r) = c_b (\ell/r_b) (r/r_b)^3$ and $\bar\phi(r) = \bar v_b - 2\log\left( r/r_b \right)$, where $c_b$ is an arbitrary constant. This corresponds to a solution with no black hole for which $\rho_{\rm eff} = - P_{\rm eff} = 3 \eta^2 M_{\rm Pl}^2 \left( 1 - 1/ (c_b \eta \ell)^2 \right) + \Lambda_4 M_{\rm Pl}^2$, where we have not assumed cancellation of the cosmological constant. If one fixes $c_b \eta \ell = 1$, then $\rho_{\rm eff} = -P_{\rm eff} = \Lambda_4 M_{\rm Pl}^2$ consistently with the 4D Einstein equations.} } 
Detailed investigation is left for a future work. 

 \subsection{Asymptotically AdS linear dilaton model}
\label{sec:Emergent_DM}

We can also define a slightly different model interpolating between AdS and the linear dilaton background. The model is defined by the bulk potential
\begin{equation}
    \bar V(\bar\phi)=\frac{1}{8}\bar W^{\prime}(\bar\phi)^2 -\frac{1}{2}\bar W(\bar \phi)^2 \,, \label{eq:V_LDA}
\end{equation}
where $\bar W(\bar\phi)=2(1+e^{\bar\phi})/\ell$ \cite{Megias:2021mgj,Megias:2021arn}. 
In the brane cosmology frame, the behavior of the effective energy term depends on the parameter $c=\exp(-\bar v_b+e^{-\bar v_b})\equiv (\eta\ell)^{-1}$. We find that $\rho_{\rm eff}$ behaves as in AdS
in the limit $c\to \infty$, and as in the linear dilaton background in the limit $c\to 0$, with
\begin{equation}
\hspace{-0.04cm}   \rho_{\rm eff} \simeq 
   \begin{cases}
    3 \eta^2 M_{\rm Pl}^2 
    \frac{r_{h}^3}{r^3_{\brane}} 
 \,\;\quad  {\rm if} ~~ c \ll 1 
 \\
 \frac{3}{\ell^2} M_{\rm Pl}^2\, \frac{r_{h}^4}{r_b^4}  \quad\quad {\rm if} ~~  c \gg 1
   \end{cases}   \,.
\end{equation}
We can recognize the  dark radiation behavior for $c\gg 1$ and the  dark matter behavior, Eq.~(\ref{eq:rhoeff}), for $c\ll 1$.

\begin{figure}[t]
\centering
\includegraphics[width=0.43\textwidth]{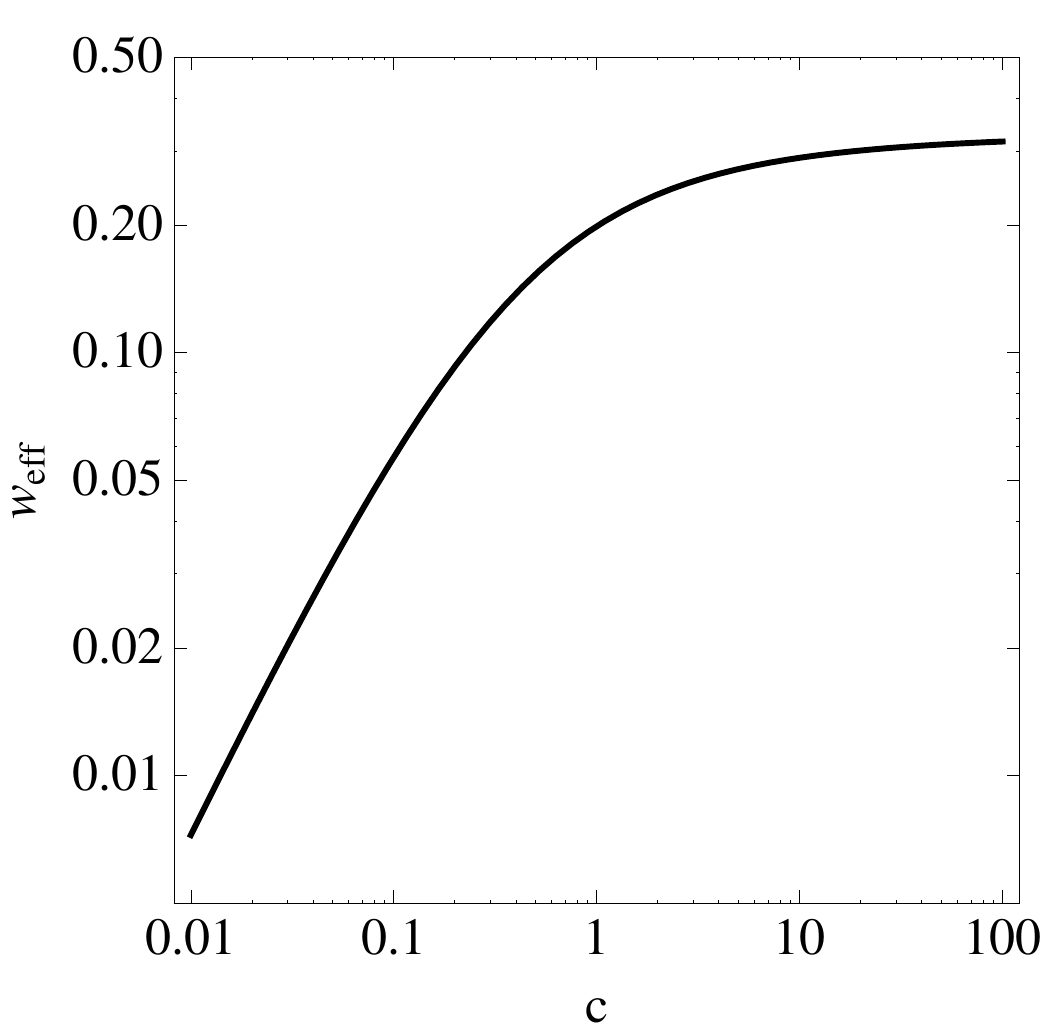}
\caption{\it Plot of the equation-of-state parameter $w_{\textrm{eff}} \equiv P_{\rm eff} / \rho_{\rm eff}$ as a function of~$c$, within the asymptotically AdS linear dilaton model.}
\label{fig:w}
\end{figure}

We confirm all these results via numerical solving of the 5D conservation equation (\ref{eq:continuity}). 
More details are given in Ref.~\cite{Fichet:2022ixi}, where we also discuss the transition region. We find that for arbitrary values of $c$, the equation of state smoothly interpolates between matter and radiation behavior, $\rho_{\rm eff}\propto a^{-3[1+w_{\rm eff}(c)]}$ with $P_{\rm eff}/\rho_{\rm eff}=w_{\rm eff}(c)$. The numerical value of the equation-of-state parameter $w_{\rm eff}(c)$ is exhibited in Fig.~\ref{fig:w} where a continuous transition appears between $w_{\rm eff} = 0$, for dark matter, and $w_{\rm eff} = 1/3$ for dark radiation.

In summary, we find that the asymptotic AdS/linear dilaton background, created by the potential in Eq.\,\eqref{eq:V_LDA}, gives rise 
to a cosmological brane-world  in which the behavior of the ``holographic'' effective energy density can range  from dark radiation to dark matter, as controlled by the $c$ parameter. 

\section{Discussion}

We now discuss a few conceptual points and relations to the literature. 

\paragraph*{Birth of the bulk black hole.}

In a typical cosmological scenario, analogously to the AdS brane-world, the bulk horizon is created by energy leaked from the brane into the conti\-nuum of bulk gravitons and other bulk fields. See~\cite{Hebecker:2001nv,Langlois:2002ke,Langlois:2003zb} for a consistent analysis in AdS, and~\cite{Fichet:2021xfn} for the rate in arbitrary background. The radiation feeds the bulk black hole, which typically grows with time. This feeding mechanism is efficient at early times while at late times, the radiation is negligible hence the  horizon does not evolve anymore. This corresponds to the low-energy regime in  our analysis. The process of dumping energy into the bulk, known since \cite{Gubser:1999vj}, is either similar or truely equivalent (via AdS/CFT) to the
 process of heating up a CFT sector (see \textit{e.g.}~\cite{Gubser:1999vj, Hebecker:2001nv, vonHarling:2012sz,vonHarling:2012sz, Brax:2019koq, Hong:2019nwd}). 

\paragraph*{What is the dark matter made of?} 
The dark matter arising in our linear dilaton  brane-world is purely made of the curvature of spacetime. However this curvature is the result of populating the  bulk with gravitons. Deep in the bulk these gravitons are strongly interacting, and their net effect  is the presence of the bulk horizon, which is seen by the brane observer.  
Since the continuum of gravitons is involved, our result shares, in a sense, some similarity with the proposal of ``continuum dark matter'' made in \cite{Csaki:2021gfm,Csaki:2021xpy,Csaki:2022lnq}. It is plausible that our analysis provides  the consistent framework needed to understand  cosmology in such models. 

\section{Prospects}

Overall, the results reported in this letter hint at an alternative view of dark matter which certainly deserves further investigation. We thus end  with a discussion of future directions.

\paragraph*{Cosmological perturbations. }
The key calculation presented in this letter shows that the linear dilaton background could explain dark matter  in the homogeneous universe. Computing perturbations and structure formation is a task beyond the present work, however the roadmap is clear: the study of cosmological perturbations in our model belongs to the realm of the fluid/gravity correspondence~\cite{Bhattacharyya:2007vjd,Hubeny:2011hd}. The dark matter of our brane-world model  amounts to  a (non-conformal) ``holographic fluid'', whose properties such as viscosities need to be carefully computed and compared to observations.

\paragraph*{Dark matter  at galactic scales. } 
Our brane-world model may explain  dark matter at cosmological scale, however no\-thing is said about galactic scales. 
To understand how  the dark matter emerging in our model behaves at galactic scales we would have  to compute less symmetric solutions of the 5D scalar-gravity system, as needed  to describe \textit{e.g.} halos. One should thus investigate  $SO(3)$-symmetric solutions, possibly assisted by matter sources on the brane.  This is left for future investigation. 

\paragraph*{Dark matter decay. } 

In analogy with AdS, the bulk black hole may in principle be able to decay via Hawking radiation into the brane, see~\cite{Rocha:2008fe,Rocha:2009xy} for an analysis in AdS. Since the bulk black hole is the origin of dark matter, Hawking decay amounts in our model to ``dark matter decay''. It would be very interesting to study this mechanism and its observational consequences, as well as its implications for holography.  We leave it as an open question to investigate.  

\paragraph*{Continuum signatures. }

In our model the graviton is accompanied by a gapped continuum that can be experimentally tested, as exemplified by the correction to the Newtonian potential Eq.\,\eqref{eq:VN_LD}. Standard Model fields can be included in the model by introducing  5D bulk fields and identifying the brane-localized modes as the Standard Model fields.   Analogously to the graviton, each Standard Model field is accompanied with a gapped continuum which has generally mild coupling to the brane. Such a setup looks typically like a dark sector~\cite{Brax:2019koq}. The  phenomenology of continuum  sectors is an active topic of investigation, see \textit{e.g.}~\cite{Katz:2015zba, Csaki:2018kxb, Lee:2018qte, Gao:2019gfw,Fichet:2019owx,Costantino:2020msc,Chaffey:2021tmj,Csaki:2021gfm,Csaki:2021xpy,Csaki:2022lnq}. The present study reinforces the motivation for such models and, in a sense, starts to explore their cosmology. 
\\

\begin{acknowledgments}
We thank Philippe Brax, Csaba Csaki and Philip Tanedo for useful discussions. The work of SF has been supported by the S\~ao Paulo Research Foundation (FAPESP) under grants \#2011/11973, \#2014/21477-2 and \#2018/11721-4 and by CAPES under grant \#88887.194785. EM would like to thank the ICTP South American Institute for Fundamental Research (SAIFR), S\~ao Paulo, Brazil, for hospitality and partial finantial support of FAPESP Grant  2016/01343-7 from Aug-Sep 2022 where part of this work was done. The work of EM is supported by the project PID2020-114767GB-I00 funded by MCIN/AEI/10.13039/501100011033, by the FEDER/Junta de Andaluc\'{\i}a-Consejer\'{\i}a de Econom\'{\i}a y Conocimiento 2014-2020 Operational Programme under Grant A-FQM-178-UGR18, and by Junta de Andaluc\'{\i}a under Grant FQM-225. The research of EM is also supported by the Ram\'on y Cajal Program of the Spanish MICIN under Grant RYC-2016-20678. The work of MQ is partly supported by Spanish MICIN under Grant PID2020-115845GB-I00, and by the Catalan Government under Grant 2021SGR00649. IFAE is partially funded by the CERCA program of the Generalitat de Catalunya.
\end{acknowledgments}

\bibliography{biblio}

\end{document}